\documentclass[oneside,a4paper]{article}

\usepackage{hyperref}
\usepackage[mathscr]{euscript}
\usepackage{verbatim}
\usepackage{hyperref}
\usepackage{tikz}
\usepackage{textcomp}
\usepackage{pgfplots}
\usepackage{enumitem}
\usepackage{amsmath}
\usepackage{float}
\usetikzlibrary{shapes,arrows}
\usetikzlibrary{shapes.misc}
\usetikzlibrary{chains,shadows.blur}
\usetikzlibrary{arrows, chains, positioning, quotes, shapes.geometric}
\pgfplotsset{width=10cm,compat=1.9}
\tikzstyle{decision} = [diamond, draw, fill=blue!20, 
    text width=4.5em, text badly centered, node distance=3cm, inner sep=0pt]
\tikzstyle{block} = [rectangle, draw, fill=blue!20, 
    text width=5em, text centered, rounded corners, minimum height=4em]
\tikzstyle{line} = [draw, -latex']
\tikzstyle{cloud} = [draw, ellipse,fill=red!20, node distance=3cm,
    minimum height=2em]

\usepackage{authblk}
\providecommand{\keywords}[1]{\textbf{\textit{Keywords:}} #1}

\title{Flexible Skyline: one query to rule them all}
\author{GIACOMO VINATI}
\affil{Politecnico di Milano\\
Milan, Italy\\
\href{mailto:giacomo.vinati@mail.polimi.it}{giacomo.vinati@mail.polimi.it} }
\date{}

\begin{document}
\maketitle
\begin{abstract}
The most common archetypes to identify relevant information in large datasets and find the best options according to some preferences or user criteria, are the top-k queries (ranking method based on a score function defined over the records attributes) and skyline queries (based on Pareto dominance of tuples). Despite their large diffusion, both approaches have their pros and cons.\\ In this survey paper, a comparison is made between these methods and the Flexible Skylines, which is a framework that combines the ranking and skyline approaches using the novel concept of $\mathscr{F}$-dominance to a set of monotone scoring functions $\mathscr{F}$.
\end{abstract}


\section{Introduction}
Handling complex problems well and considering multiple criteria explicitly leads to more informed and better decisions. 
Multiple criteria decision making, deals with mathematical optimization problems ($\mathit{min}$, $\mathit{max}$ of a function) involving more than one objective function to be optimized simultaneously and evaluating multiple conflicting criteria in decision making, where optimal decisions need to be taken in the presence of trade-offs between two or more colliding objectives.\\
Now, considering dataset which consists of many objects and each of them is described by some features or attributes: out of this data , most of the times, is necessary to find the best options according to some preferences to some user criteria and the aim of simultaneously optimize two or more quality criteria is known as multi-objective optimization.\\ It’s used for mainly two different data mining scenarios:
the first one involves predictive tasks, such as classification, regression and the second one attribute \& feature selection.
Three are the mainly types of approaches to cope with multi-objective optimization:
\begin{itemize}
  \item The lexicographic approach (not debated in this paper).
  \item The ranking queries (Top-k) approach that depend on the choice of weight in a scoring function for ranking tuples.
  \item The skyline approach that uses the Pareto dominance to retrieve the best tuples that are not dominated by any other tuple.
\end{itemize} 
 In the next sections is presented a description of each methods with their pros and cons and a comparison with the newest Flexible Skylines approach.
\section{Ranking queries}
\label{sec:rank}One of the most used approach for multi-objective optimization consists in transforming it into a single-objective problem, assigning a numerical weight to each attribute (chosen from the dataset $\mathit{R}$) and combine the values in a single one  by adding or multiplying them.
One common way to identify the top-k objects, is by using a scoring function $\mathscr{S}$ and below is described its general formula \cite{12}:
\begin{equation}
{S = w_{1}\times a_{1} + w_{2}\times a_{2} + ... + w_{n}\times a_{n}}\label{eq:1}
\end{equation}
where \(w_{i}=1,...n \) represents the weight associated to each attribute \(a_{i}=1,...n \).This approach is one of the most used due to the fact that is possible retrieve the relative importance of attributes, control the cardinality of the result and possibly reach a compromise among different attributes \cite{3}.\\The other side of the coin is that ranking queries depends on the choice of the weight in the scoring function and this may require to repeat the task multiple times in order to "predict" the effect on ranking by changing one or more parameters of the scoring function $\mathit{S}$\eqref{eq:1}.
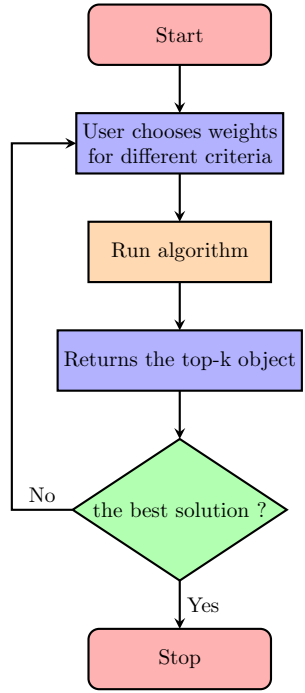
\begin{figure}[H]
\centering
    \begin{tikzpicture}[scale=0.8, transform shape,
    node distance = 8mm and 20mm,
      start chain = going below,
     arrow/.style = {thick,-stealth},
      base/.style = {
            draw, thick, 
            minimum width=30mm, minimum height=10mm, align=center,
            inner sep=1mm, outer sep=0mm,
            on chain, join=by arrow},
  decision/.style = {diamond, base,
            aspect=1.5, inner xsep=0mm,fill=green!30},
   process/.style = {rectangle, base, fill=orange!30},
   input/.style = {rectangle, base, fill=blue!30},
 startstop/.style = {rectangle, rounded corners, base,fill=red!30},
                        ]
\node (start)   [startstop] {Start};
\node (setup)   [input]   {User chooses weights \\ for different criteria};
\node (compute) [process]   {Run algorithm};
\node (compute2) [input]   {Returns the top-k object};
\node (test)    [decision]  {the best solution ?};
\node (displace)[startstop]   {Stop};

\draw [arrow] (test.west) to["No" '] + (-1,0) |-  (setup.west);
\path   (test) to["Yes"] (displace);

\end{tikzpicture}
\caption{Top-k query Flowchart} \label{fig:flowchart1}
\end{figure}
Another downside is that this approach suffers from scalability problems with respect to the database size and the number of features \cite{9}.These approach touches different fields such as caching with the purpose of increasing data retrieval performance by reducing the need to access the next, slower tier of storage. It's also used in multimedia search, e-commerce, metasearch engines for the efficient combining of rankings from different search engines and many other fields (ML, Recommended systems).\\Top-k processing techniques can be classified into different categories based on query model, data access methods, implementation level, data \& query uncertainty and ranking function.\\The latter can also be divided in three sub-categories: the first one, which includes the majority of current techniques, assumes the usage of a monotone ranking functions; the second category make use of generic ranking functions \cite{19} and the third one leaves the ranking function unspecified(this leads to skyline queries) \cite{9,14}.\par\null\par\bigskip\emph{Example 1.1}\label{sec:ex11} Considering a user interested in finding a particular PC model in different stores where the cost of the Pc and the distance of the store is minimized. This is an example of a dataset with just two dimensions: price and distance. The preference criterion in the scoring function tells exactly how the user want to rank the object which is really a very important part of the problem.In a first run the user set the score function(e.g.  \(Score=distance+3*price \)) and result show the \(t_{1}\) tuple as the top query. Then, with the same data set, if the user computes the global score of these Pc's with a different criterion (for example user weights seven times the distance as much as the price \(Score=7*distance+price \) ) then basically that means that user explores this data set according to some other direction, so in this case the best Pc for the user is another one (\(t_{3} \)).
 \begin{figure}[H]
 \centering
\begin{tikzpicture}[font=\sffamily,
        myCircle/.style={circle, draw, minimum size=4.5mm, inner sep=0pt, thick},
        rectCircle/.style={myCircle, fill=red},
            ]
\begin{axis}[
    title={},
    xlabel={distance},
    ylabel={price},
    xmin=0, xmax=50,
    ymin=0, ymax=50,
    xtick={0,10,20,30,40,50},
    ytick={0,10,20,30,40,50},
    ymajorgrids=false,
]      
 \addplot [red,thick,domain=0:50, samples=100]{-0.333*x+21};
 \addplot [blue,thick,domain=0:50, samples=100]{-7*x+104};
 \addlegendentry{\scriptsize\(Score=distance+3*price\)}
 \addlegendentry{\scriptsize\(Score=7*distance+price \)}\textbf{}
\end{axis}

    \node [rectCircle, draw] (c1) at (6,1.3) {};
    \node [myCircle, draw] (c2) at (4.2,2.7) {};

    \node [rectCircle, draw, fill = blue] (c4) at (1.4,6.4) {};

    \node [myCircle] (c5) at (3.5,4.7){};
    \node [myCircle] (c6) at (5.1,3.2){};
    \node [myCircle] (c7) at (7.1,1.8){};
    \node [myCircle] (c9) at (2.3,6.7){};
    \node [myCircle] (c10) at (2.15,4.1){};
    
    \draw (c1) node [below=2mm] {$t_1$};
    \draw (c2) node [below=2mm] {$t_2$};

    \draw (c4) node [left=2mm] {$t_3$};
    \draw (c5) node [below=2mm] {$t_4$};
    \draw (c6) node [below=2mm] {$t_8$};
    \draw (c7) node [below=2mm] {$t_6$};        
    \draw (c9) node [below=2mm] {$t_7$};
    \draw (c10) node [below=2mm] {$t_5$};

\end{tikzpicture}
\caption{example 1.1 Top-k query} \label{fig:figfig1}
\end{figure}
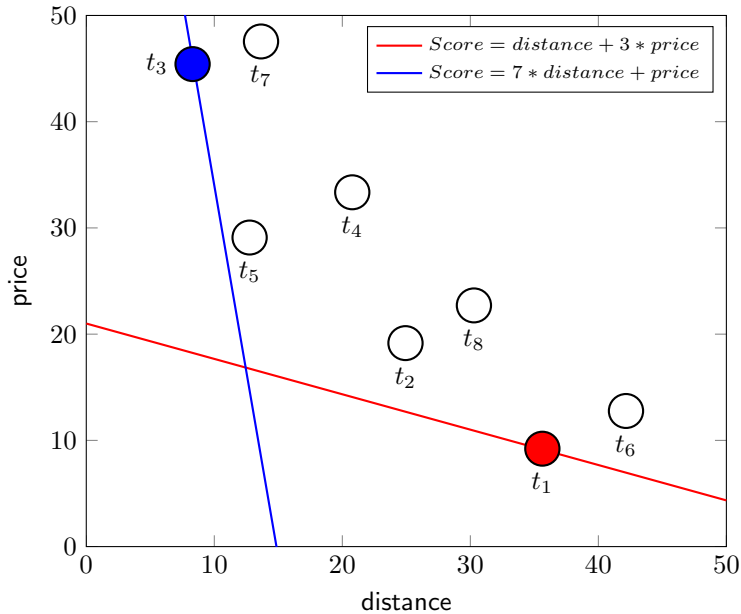
As shown in the figure (\ref{fig:figfig1}), if the scoring function is changed, also the preferences are changed then also the result is different and this is a rather important topic because there are datasets and problems in which the preference criterion is really crucial and even very small changes in that, might determine big changes in the result. 
\newline So the result of the query heavily depends on how the user preferences are translated in the choice of weights in the scoring function. This is a very hard task due to the fact that it's difficult to predict the effect of changing the wights on the final ranking result and thus fail to offer an overall view of the dataset.\\ Also the user cannot quantify with absolute precision the relative importance of the different attributes \cite{1,4,10}.

\section{Skyline queries}
\begin{figure}[H]
\centering
\begin{tikzpicture}[scale=0.8, transform shape,
    node distance = 8mm and 20mm,
      start chain = going below,
     arrow/.style = {thick,-stealth},
      base/.style = {
            draw, thick, 
            minimum width=30mm, minimum height=10mm, align=center,
            inner sep=1mm, outer sep=0mm,
            on chain, join=by arrow},
  decision/.style = {diamond, base,
            aspect=1.5, inner xsep=0mm,fill=green!30},
   process/.style = {rectangle, base, fill=orange!30},
   input/.style = {rectangle, base, fill=blue!30},
 startstop/.style = {rectangle, rounded corners, base,fill=red!30},
                        ]
\node (start)   [startstop] {Start};
\node (compute) [process]   {Run the algorithm};
\node (compute2) [input]   {Returns all non-dominated \\tuples/solutions};
\node (test)    [process]  {User chooses the best solution \\according to his/her \\preferences a posteriori};
\node (displace)[startstop]   {Stop};

\end{tikzpicture}
\caption{Skyline query Flowchart} \label{fig:flowchart2}
\end{figure}
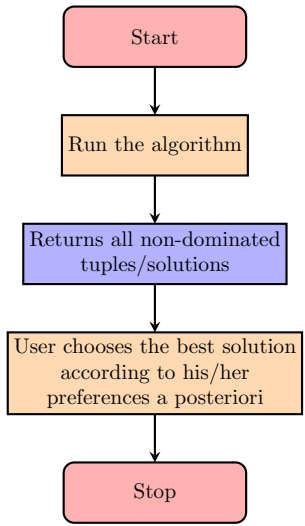

The objective of this approach is to return all the non-dominated tuples.This is also known as a maximal vector problem \cite{14} which that aims to identify the maximals over a set of vectors, or more specifically, find the subset of vectors where each one is not dominated by any other vector in the set \cite{13}.\\With this approach, instead of transforming a multi-objective problem to a single one, a multi-objective algorithm is used for solving the original multi-objective problem. 
This is also called the Pareto dominance approach \cite{5}, where in a dataset R, a tuple $t$ is said to dominate a tuple $s$ if and only if exists at least an attribute of $t$ that is greater then the same attribute of $s$ and if all the attributes of $t$ are greater than or equal to the respective attributes of $s$. In a more formal approach \cite{12}:
\begin{equation}
{SKY(R) =\{t\in R \mid \forall s  \in R, s\neq t, \exists a_{i} ( t(a_{i}) >s(a_{i}) \wedge \forall a_{k}, k=1,...n, t(a_{k}) \geq s(a_{k}))\}}
\end{equation}
With respect to ranking queries, the skyline approach has the advantage of simplicity and easy of use, due to the fact that there is no weight parameter to be set by the user \cite{12} and even if the scoring function is not known, the highest scoring tuple is guaranteed to be in the skyline.
\\Even if there is high probability that the query may retrieve too many object as result (cardinality issue and lack of customization in controlling the number of outputs), this gives to the user different solutions that allows him/her to analyze different trade-offs (Figure~\ref{fig:flowchart2}) associated with different criteria and choose the best solution a posteriori \cite{4,12}.
\emph{Example 1.2} Taking the \hyperref[sec:ex11]{same example used for top-k queries}, 
there is the same dataset with normalized attribute values, but instead of deciding which one is the best object, which one is the second best and so on and so forth, the query returns which ones are those that might be the best options for at least one preference criterion and in particular the black points in the skyline area are those that are potentially the best ones for some criteria and that are not dominated by any other point.
\begin{figure}[H]
\centering
\begin{tikzpicture}[scale=0.95, transform shape, font=\sffamily,
        myCircle/.style={circle, draw, minimum size=5mm, inner sep=0pt, thick},
        rectCircle/.style={myCircle, fill=black},
        blankCircle/.style={circle,draw, minimum size = 0.5mm, color = white},
        cross/.style={cross out, minimum size=1.5mm, draw, inner sep=0pt, solid}
            ]
\begin{axis}[
    title={},
    xlabel={\large distance},
    ylabel={\large price},
    xmin=0, xmax=1,
    ymin=0, ymax=1,
    xtick={0.0,0.2,0.4,0.6,0.8,1},
    ytick={0.0,0.2,0.4,0.6,0.8,1},
    ymajorgrids=false,
    no markers,
]      

  \node[] at (axis cs: 0.2,0.3) {\large skyline};
     \node[] at (axis cs: 0.7,0.7) {\large non-skyline};

\end{axis}

    \node [rectCircle, draw] (t1) at (6,1.2) {};
    \node [rectCircle, draw] (t2) at (4.1,2.8) {};
    \node [rectCircle, draw] (t5) at (2.2,4) {};
    \node [rectCircle, draw] (t3) at (1.5,6.3) {};

    \node [myCircle] (t6) at (3.5,4.4){};
    \node [myCircle] (t7) at (4.9,3.2){};
    \node [myCircle] (t8) at (6.8,1.8){};
    \node [myCircle] (t9) at (2.3,6.65){};
    
    \draw (t1) node [below=2mm] {$t_1$};
    \draw (t2) node [below=2mm] {$t_2$}; 
    \draw (t3) node [below=2mm] {$t_3$};
    \draw (t7) node [right=2mm] {$t_8$};
    \draw (t5) node [below=2mm] {$t_5$}; 
    \draw (t6) node [right=2mm] {$t_4$};
    \draw (t8) node [right=2mm] {$t_6$};
    \draw (t9) node [right=2mm] {$t_7$};
    \coordinate (orig) at (0,0);

\draw [](1.5,7) -- (t3) -- (2.2,6.3) -- (t5) -- (4.1,4) -- (t2) -- (6,2.8) -- (t1) -- (8.4,1.2);

\end{tikzpicture}
\caption{example 1.2 Skyline query} \label{fig:figfig2}
\end{figure}
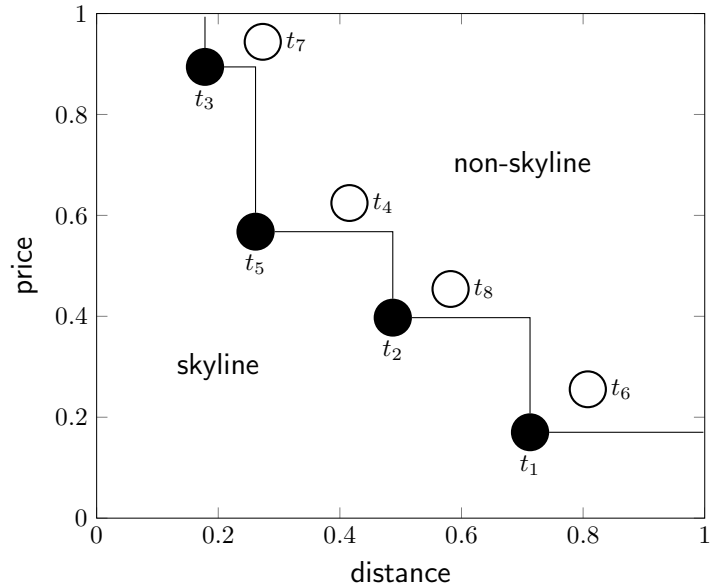
In the example (Figure~\ref{fig:figfig2}) if the user gives a lot of importance to distance, then the black point in the left upper part would be the best option although it has a high price.
The lower-right black point is instead the best option if user gives a lot of importance to price although it has a high distance.\\White ones are points that are certainly not the best options because they are dominated by at least one of the black points, so they can never be the best option no matter what preference criterion user chooses and this kind of query is called a skyline query because it kind of offers the peaks in a dataset that give user a nice overview of what is potentially interesting in a dataset.

\section{Restricted Skylines}
As said in the previous sections, ranking and skyline queries have their own weaknesses: first of all, Top-k retrieves the top best objects but the choice of the right weights in the scoring function is a complex problem for the user, cause each weight is usually empirically determined. Additionally, the user must pay attention in comparing, in the scoring function, only commensurable criteria in order to avoid meaningless results or quantities \cite{12}. \newline Skyline approach instead, suffers from cardinality issue, such that the user itself cannot control the output dimension and his/her preferences cannot be satisfied due to the fact that this approach returns a set a non-dominated solutions, whereas in practice the user will often use a single best one.\\A novel skyline approach for multi-objective optimization, called Flexible Skylines (aka F-skylines) aims to combine pros of ranking and skyline queries and capture the user preferences or model priority criteria by means of arbitrary constraints on the space of the weights without the hard task of specifying weights in a score function.\\This original framework is aware of users' criteria by means of constraints on the weights used in a scoring function \cite{1} and for this purpose the concept of $\mathscr{F}$-dominance is lead.
\\A tuple $\mathit{t}$ is said to $\mathscr{F}$-dominating a tuple $\mathit{s}$ when $\mathit{t}$ is always better than or equal to $\mathit{s}$ with respect to all the monotone scoring functions that belong to a limited set $\mathscr{F}$ of monotone scoring functions \cite{1,2,3}.
More generally, if the normal skyline approach consider the dominance (maximal vector problem or Pareto dominance)\cite{14} over a set $\mathscr{M}$ of monotone scoring functions, the F-skylines apply the same method but on a subset $\mathscr{F}$ of $\mathscr{M}$.
\\Taking into account a relational schema $\mathscr{R}$ with $\mathscr{A}$ attributes $\mathscr{R}(A_{1},...A_{d})$ with $\mathit{d}\geq1$ with normalized attributes values in range \([0,1]\) and the instance \(\mathit{r} \in \mathscr{R}\), skylines and F-skylines are described in a more formal approach as follows:
\begin{equation}
{SKY(r) = \{ t \in r \mid \exists f \in \mathscr{M}. \forall s \in r. s\neq t \rightarrow f(t) < f(s)\}}\label{eq:3}
\end{equation}
\begin{equation}
{F-SKY(r) = \{ t \in r \mid \exists f \in \mathscr{F\subseteq M}. \forall s \in r. s\neq t \rightarrow f(t) \leq f(s)\}}\label{eq:4}
\end{equation}
 The  $\mathscr{F}$-dominance concept in notation \eqref{eq:4} can also be written as \( t\prec_\mathscr{F} s\).
\\The F-skylines also introduce two flexible operators which both return a subset of the skyline: $\mathit{ND}$ and $\mathit{PO}$ operators.
The first refers to the set of non-$\mathscr{F}$-dominated tuples and the second one to the potentially optimal tuples \cite{1,2,3}.\\
As said in \cite{1}, for any set $\mathscr{F}$ of monotone scoring functions, $\mathit{ND}$ is a subset of the skyline and $\mathit{PO}$ is a subset of ND and both parameters reduce the size of the final result (cardinality).\\
The high flexibility of this operators allows to solve the multi-objective function problems, retrieving the top one query and also constraint the scoring function and being very effective on large datasets.
\begin{figure}[H]
\centering
    \begin{tikzpicture}[node distance=2cm, scale=0.75, transform shape]
    \tikzstyle{startstop} = [rectangle, rounded corners, minimum width=3cm, minimum height=1cm,text centered, draw=black, fill=red!30]
    \tikzstyle{io} = [rectangle, minimum width=3cm, minimum height=1cm, text centered, draw=black, fill=blue!30]
    \tikzstyle{process} = [rectangle, minimum width=3cm, minimum height=1cm, text centered, draw=black, fill=orange!30]
    \tikzstyle{decision} = [diamond, minimum width=3cm, minimum height=1cm, text centered, draw=black, fill=green!30]
    \tikzstyle{arrow} = [thick,->,>=stealth]
    \node (start) [startstop] {Start};
    \node (in1) [io, below of=start] {User choose constraints on weights};
    \node (dec1) [decision, below of=in1, yshift=-1cm] {need precision?};
    \node (pro2a) [process, below of=dec1, yshift=-1.5cm] {Run ND algorithm};
    \node (pro2b) [process, right of=dec1, xshift=3cm] {Run PO algorithm};
    \node (out1) [io, below of=pro2a] {Returns the ND tuples};
     \node (out2) [io, below of=pro2b] {Returns the PO tuples};
    \node (dec2) [decision, below of=out1, yshift=-1cm] {the best solutions?};
    \node (stop) [startstop, below of=dec2, yshift=-1cm] {Stop};

    \draw [arrow] (start) -- (in1);
    \draw [arrow] (in1) -- (dec1);
    \draw [arrow] (dec1) -- node[anchor=east] {no} (pro2a);
    \draw [arrow] (dec1) -- node[anchor=south] {yes} (pro2b);
    \draw [arrow] (pro2a) -- (out1);
    \draw [arrow] (pro2b) -- (out2);
    \draw [arrow] (out1) -- (dec2);
    \draw [arrow] (out2) |- (dec2);
    \draw [arrow] (dec2) -- node[anchor=east] {yes} (stop);
    \draw [arrow] (dec2.west) to["no" '] + (-2,0) |-  (in1.west);
    
\end{tikzpicture}
\caption{F-skyline query Flowchart} \label{fig:flowchart3}
\end{figure}
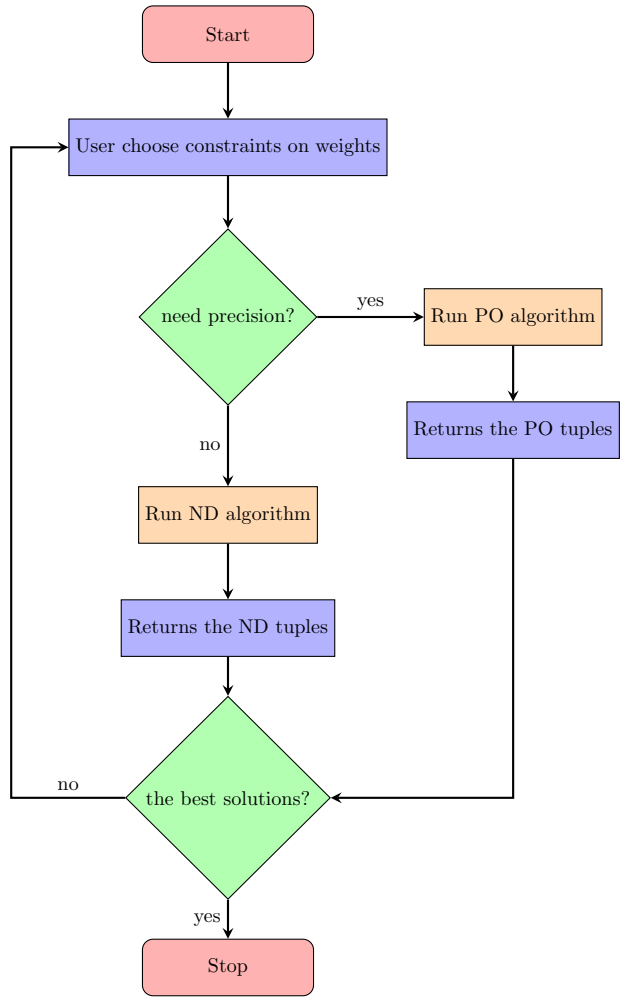
As shown in the flowchart (\ref{fig:flowchart3}), the result of the query may be quite different, according to choice of the constraints and the desired operator: select the ND will return a set of non-dominated tuples; however It will be more precise than the classic skyline approach due to the fact that the result set is a subset of the latter(SKY).\\Selecting the PO instead, will return as result a subset of ND, containing only the potentially optimal solutions (tuples), that are strictly better than all the others. This is true for any set $\mathscr{F}$ of monotone scoring functions \cite{1}.\par\null\par\bigskip\emph{Example 1.3} Using F-skylines framework over the same dataset used for the previous examples, the user's preference can be translated into a constraint on weights of the scoring function \cite{1,2} (e.g. set of monotone scoring function \( \mathscr{F} =\{ \mathit{w_{p}}*price + \mathit{w_{d}*distance} \mid w_{p} \geq w_{d}\} \) ) where the price affects more than distance (e.g. \(w_{p}=0.9,w_{d} = 0.1 \)).
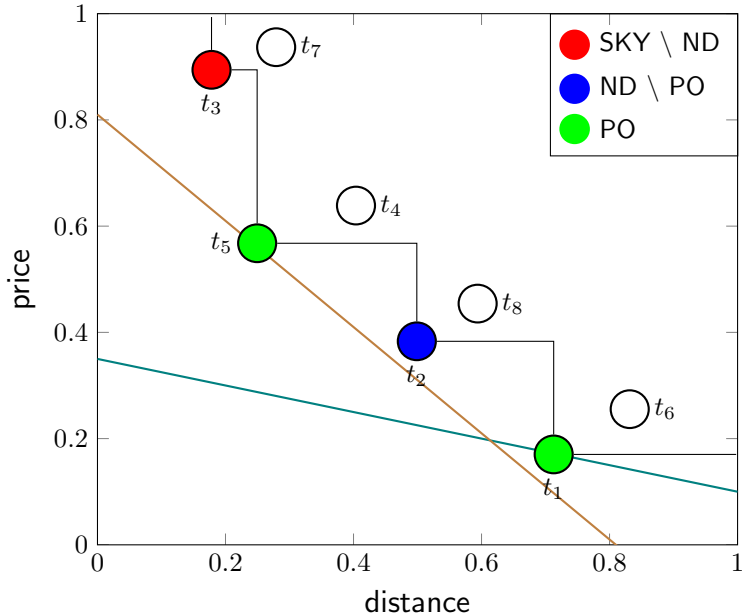
\begin{figure}[H]
\centering
\begin{tikzpicture}[font=\sffamily,
        myCircle/.style={circle, draw, minimum size=5mm, inner sep=0pt, thick},
        rectCircle/.style={myCircle, fill=black},
        blankCircle/.style={circle,draw, minimum size = 0.5mm, color = white},
        cross/.style={cross out, minimum size=1.5mm, draw, inner sep=0pt, solid},
        blacknode/.style={shape=circle, draw=black, line width=2},
        bluenode/.style={shape=circle, draw=blue, line width=2},
        greennode/.style={shape=circle, draw=green, line width=2},
        rednode/.style={shape=circle, draw=red, line width=2}
            ]
\begin{axis}[
    title={},
    xlabel={\large distance},
    ylabel={\large price},
    xmin=0, xmax=1,
    ymin=0, ymax=1,
    xtick={0.0,0.2,0.4,0.6,0.8,1},
    ytick={0.0,0.2,0.4,0.6,0.8,1},
    ymajorgrids=false,
    no markers,
]      
 \addplot [teal,thick,domain=0:1, samples=100]{-0.25*x+0.35};
 \addplot [brown,thick,domain=0:1, samples=100]{-1*x+0.81};

\end{axis}

    \node [rectCircle, draw,fill = green] (t1) at (6,1.2) {};
    \node [rectCircle, draw,fill = blue] (t2) at (4.2,2.7) {};
    \node [rectCircle, draw,fill = green] (t5) at (2.1,4) {};
    \node [rectCircle, draw, fill = red] (t3) at (1.5,6.3) {};

    \node [myCircle] (c0) at (3.4,4.5){};
    \node [myCircle] (c1) at (7,1.8){};
    \node [myCircle] (c8) at (2.35,6.6){};
    \node [myCircle] (c9) at (5,3.2){};

    \draw (t1) node [below=2mm] {$t_1$};
    \draw (t2) node [below=2mm] {$t_2$};
    \draw (t3) node [below=2mm] {$t_3$};
    \draw (t5) node [left=2mm] {$t_5$};
    
    \draw (c0) node [right=2mm] {$t_4$};
    \draw (c1) node [right=2mm] {$t_6$};
    \draw (c8) node [right=2mm] {$t_7$};
    \draw (c9) node [right=2mm] {$t_8$};

    \coordinate (orig) at (0,0);

\draw [](1.5,7) -- (t3) -- (2.1,6.3) -- (t5) -- (4.2,4) -- (t2) -- (6,2.7) -- (t1) -- (8.4,1.2);
    
    \matrix [draw, below right] at (5.95,7.05) {
  \node [rednode,fill=red, label=right:SKY \(\setminus\) ND] {}; \\
  \node [bluenode,fill=blue,label=right:ND \(\setminus \) PO] {}; \\
  \node [greennode,fill=green,label=right:PO] {}; \\
};

\end{tikzpicture}
\caption{example 1.3 \(\mathit{F}\)-Skyline query} \label{fig:figfig3}
\end{figure}
As shown in the figure (\ref{fig:figfig3}), the F-Skyline, instead of showing only the non-dominated tuples, as said before, the user can translate the preference (e.g. in price) into a constraint on weight of the scoring function. In such case $t_{1}$,$t_{2}$,$t_{5}$ are the non-dominated tuples (Pc's) and $t_{3}$ belongs to the SKY even if It's F-dominated by $t_{5}$, since It has an higher price.
\\Between the potentially optimal $\mathit{PO}$ $t_{1}$,$t_{5}$, the first one would be the best choice (the teal colored line) if the user gave a lot of importance to the price (i.e.\(w_{p}=0.8, w_{d}=0.2 \)).On the other hand, $t_{5}$ would be the best option  if the user gave the same importance (the brown colored line) to price and distance (i.e.\(w_{p}=0.5, w_{d}=0.5 \)). Tuple $t_{2}$ is $\mathit{ND}$ but not $\mathit{PO}$ since there is no combination of weights that satisfies the initial constraint \( w_{p} \geq w_{d} \) and that makes it the best choice over any other tuple.

\subsection{Comparison}
General skylines consist in retrieving the maximals or admissible points (maximal vector problem) of a monotone scoring function and even if it's unknown by the user, the set of non-dominated tuple is guaranteed to be in the skyline.\\
The general issues with this approach are essentially three:
\begin{enumerate}
\item It's an high cost operation.
\item The cardinality of the output is out of control for the user and this might be useless or not informative enough for choosing the best final option.
\item Also, the limit of this approach is that it assigns the same importance to all attributes and can't represent user's preferences.
\end{enumerate}
F-skyline approach instead, uses a scoring function on a subset of a monotone scoring functions, and constraining the weights, allows the preference elicitation of the user \cite{7}.The F-skyline operator ($\mathit{ND}$ \& $\mathit{PO}$) are very effective in focusing on the tuple of interest for the user, and can lead to a considerably smaller result (reducing the cardinality problem of standard skylines) and the more the constraints are restrictive, the more the results of both $\mathit{ND}$ and $\mathit{PO}$ reduce \cite{1}.\\A possible common problem for both approaches, is the normalization of different quality attributes: choosing the right method (min-max,z-score, decimal scaling, etc...) can be affected by an inductive bias that would prioritize one solution (or interpretation) over another, independent of the observed data.Top-k approach instead, solves a multi-objective problem, transforming it in a single-objective one and using a scoring function to retrieve the top-k objects, turning out to be more representative of user preferences with respect to standard skyline queries and even if It has the advantage of conceptual simplicity and easy of use, it can lead to uncertain results, due to the fact that the user has to specify exact weight values without knowing how they condition the final output.Sometimes, the uncertainty is upstream represented by the very own values of the dataset due to human/sensor error and this is a problem for determining the right ordering of Top-k result \cite{10,11}.\\F-skyline approach instead, solves the problem of the empiric estimation of weights, by constraining them to a subset of a monotonic scoring function $\mathscr{F}$.Regarding the uncertainty of the data, the latter query approach can deal with this by learning weights via crowdsourcing tasks and reducing the uncertainty, modeling it as constraint on wights of the scoring function \cite{11}.\\A possible common problem for both approaches, might be the right choice of commensurable criteria that have to be added/subtracted in the scoring function, cause comparing different attributes can produce a meaningless quantity that is essentially useless for the final user's choice.

\subsection{Area of applicability}

Possible application areas of Flexible skylines are where the ranking problem is relevant and
the following list displays only few of the dozens possible different field of applicability, most of which are discussed in \cite{15,18}. 
\begin{enumerate}
  \item Search engines: for example if a user want to book an hotel or a restaurant or search for a new car, this approach can express several different criteria that match user's preferences (multiple criteria $\rightarrow$ multi-objective optimization), so the goal of the search engine is to propose to the user those options that best match his/her criteria.
  This approach is also found in e-commerce where the user search for different articles based on his/her preferences.
  \item Machine-learned ranking (MLR) or learning to rank, where F-skylines could improve the quality of ranking models for information retrieval.
  \item NLP with Text mining \& Sentiment Analysis: text pre-processing and feature selection are crucial steps in text mining and F-skyline could improve text classification accuracy, giving a competitive advantage in business over competitors.
  \item Recommender systems: algorithms that mimic the psychology and personality of humans, in order to predict their needs and desires.\\ Retrieve different user's preferences on his past searches or usages of the service, aggregate different recommendations and generate a global one that combines different criteria.   
  \item Feature Selection: when the number of features in the dataset is very large, It may contain a lot of noisy data and the more relevant and sensible features user select for the model creation, the faster is the output and the better is the accuracy of the model and reasons for using F-skyline approach in feature selection are\cite{20}:
  \begin{enumerate}
   \item Enable the ML algorithm to train faster.
    \item Reduce the complexity of a model and makes it easier to interpret.
    \item Improving the accuracy of a model.
    \item Reduce overfitting.
  \end{enumerate}
   
\end{enumerate}
  Some more specific examples could be:
  \begin{enumerate}
      \item Labeling similar images that later are fed into Neural Network for images recognition (g.e. useful for self driving vehicles, faster and more accurate facial recognition,...).
      \item Credit institutions that use users' criteria for finding the best candidates,the ones that would potentially get a loan and the ones which wouldn't.
      \item Better ranking of players in lobbies in online video games matches.
      \item Improve anti-cheat systems always in online videogames.
      \item Identify genes relevant to disease in Bioinformatics.
      \item Drug discovery: identify the most promising structure in a chemical library.
      \item Detecting Malwares.
      \item Improving sound classification in Automatic Speech Recognition applications (ASR).
      \item Rank all Tweets, Instagram \& Facebook posts and improve the timeline composition in reverse chronological order.
\end{enumerate}

\section{Conclusion}
Top-k, skyline and F-skyline approaches has been surveyed: starting from multi-objective optimization and the maximal vector problem (or Pareto dominance), the description of each method has been provided with its respective pros and cons.
By comparing them, comes to light how Flexible skylines approach overcomes the weaknesses of Top-k queries, giving the possibility of constraining the weights of the scoring function and of normal Skyline using the ND \& PO operators and reducing the cardinality of the result, taking into account the user's preferences and criteria.

\bibliographystyle{unsrt}
\bibliography{refs.bib}
\end{document}